\documentstyle{article}
\begin{document}

\title{ Quantum Hydrodynamics of Fermi Fluids }
\author{ Girish S. Setlur \\ The Institute of Mathematical Sciences
 \\ Taramani, Chennai 600113 }
 
\maketitle                 

\begin{abstract}
 It is shown that gauge theories with fermions 
 are most naturally studied via a polar decomposition of the field variable. 
 This is the fermionic analog of the preprint
 cond-mat/0210673. The hope is that these two put together will
 enable the treatment of neutral nonrelativisitc 
 matter composed of electrons and nuclei in a nonperturbative manner with
 nuclei and electrons treated on an equal footing.
 We recast the electron-phonon (superconductivity) 
  problem in the hydrodynamic language
 and indicate how it is solved. In particular we focus on
 the a.c. conductivity.
\end{abstract}

\section{Introduction}

 {\it{ Read My Lips ! }}
 
 \hspace{2.0in} George Bush Sr., 1993 Campaign.

 \vspace{0.2in}

 The program of quantizing hydrodynamics has a long and distinguished history. 
 Landau \cite{Landau} and his students
 were among the first to attempt this. Later
 on Sunakawa et.al. \cite{Sunakawa} and others - notably Rajagopal and Grest
 \cite{Raja} took this program further. Dashen, Sharp, Menikoff and
 Goldin\cite{Sharp} in the seventies introduced many of these ideas.
  Recently, Jackiw and collaborators\cite{Jackiw}
  have revived interest in this approach in the context of relativistic quarks.
 In our earlier work, we introduced the DPVA for fermions\cite{Setlur1}.
 We also note that Rajagopal and Grest\cite{Raja} had already in the
 seventies pointed out the need
 for having a nonzero-phase functional found in the DPVA.
 In our earlier work\cite{Setlur1} we
 made a first pass at computing the phase functional.
 This attempt yielded an answer that in retrospect is quite wrong.
 Upon closer examination 
 the $ U_{0}( {\bf{q}} ) $ of our earlier work\cite{Setlur1} is imaginary
 when it was postulated to be real(for small $ {\bf{q}} $). 
 So far the author has avoided this issue by taking refuge under the
 the sea-boson approach that enables us to derive the momentum distribution,
 anomalous exponents, quasiparticle residue and so on without yielding the
 full dynamical propagator which is of interest
 only because it contains information
 about quantities just mentioned. If one is able to compute them without
 having to compute the full propagator so much the better.
 However, there are physical problems
 in which the full propagator is important.
 The X-ray edge problem\cite{Mahan} is one such.
 In fact we tried using the DPVA to
 compute the X-ray edge spectra in a
 preprint\cite{preprint} and found that we obtain the
 right answers in one dimension but the answers in higher dimensions were
 inconsistent with Mahan's exact results\cite{Mahan}. 
 Thus we must now face up to the harsh reality and try and address this 
 (hopefully last) hurdle. This article is the fermionic analog of the 
 preprint cond-mat/0210673. In an earlier preprint, after much reflection,
 we chose to dismiss the approach that only uses the hydrodynamic
 variables namely the density and its conjugate as `myopic' 
 (mypoic bosonization). This is beacause a hamiltonian formulation in terms
 of the hydrodynamic variables is unable to distinguish between fermions
 and bosons. We have to further decompose these variables in terms of linear
 combination of oscillators in order to distinguish between the two statistics.
 However, the sea-boson approach is not without its share of problems. For one
 it does not generalise to finite temperatures easily. Also the full dynamical
 propagator is not reducible to quadratures due to a technical difficulty.
 Both these problems may be resolved in an approach that incorporates
 only the hydrodynamical variables.  
 We show in this preprint, that the path integral approach is 
 an avenue to distinguish between the statistics when using only the 
 hydrodynamical variables.  
 
 The author had this idea in 1993 but conversations with various knowledgeable
 people suggested that this approach that only uses hydrodynamic variables
 is not going to work out for fermions, since one has to take into account
 the extended nature of the Fermi surface. However, this idea seems
 too important to pass up. In particular, the natural manner in which 
 gauge theory may be studied in this approach makes this effort 
 for fermions worthwhile and urgent.

\section{ The Field Operator }

 In our earlier work\cite{Setlur1},
 we showed how the field operator may be expressed in
 terms of currents and densities. We reproduce the formula here.
\begin{equation}
\psi({\bf{x}}) = e^{-i \Pi({\bf{x}}) } e^{i \Phi([\rho];{\bf{x}}) }
 \sqrt{ \rho({\bf{x}}) }
\end{equation}
The conjugate $ \Pi $ obeys canonical commutation rules.
\begin{equation}
[\Pi({\bf{x}}), \Pi({\bf{x}}^{'})] = 0
\end{equation}
\begin{equation}
[\Pi({\bf{x}}), \rho({\bf{x}}^{'})] =
 i \mbox{   }\delta^{d}({\bf{x}}-{\bf{x}}^{'})
\end{equation}
\begin{equation}
[\rho({\bf{x}}), \rho({\bf{x}}^{'})] = 0
\end{equation}
 There are some technical difficulties associated with the fact that
 $ \Pi $ is not strictly self-adjoint, but we shall operate in
 the high enough density limit where we may assume that we expand around
 a nonzero mean for the density and this means that we may choose $ \Pi $ 
 to be self-adjoint.
 The conjugate $ \Pi $ may be related to the current as follows.
\begin{equation}
{\bf{J}}({\bf{x}}) = -\rho({\bf{x}})\nabla \Pi({\bf{x}})
 + \rho({\bf{x}})  \nabla \Phi([\rho];{\bf{x}}) -
\rho({\bf{x}}) [-i\Phi([\rho];{\bf{x}}),\nabla \Pi({\bf{x}})] 
\label{CURR}
\end{equation}
 The phase functional $ \Phi $ determines the statistics of the field $ \psi $.
 As shown earlier\cite{Setlur1} $ \Phi $ obeys a recursion relation 
 as depicted below. It must be pointed out that the need for having
 a nonzero $ \Phi $ for fermions was appreciated by
 Rajagopal and Grest\cite{Raja} way back in the seventies. However,
 the constraint below brought about by the imposition of Fermi statistics
 on $ \psi $ was probably first shown in our earlier work\cite{Setlur1}.
\[
\Phi([\{\rho({\bf{y}}) - \delta^{d}({\bf{y}}-{\bf{x}}^{'})\}];{\bf{x}})
- \Phi([\rho];{\bf{x}})
\]
\begin{equation}
- \Phi([\{\rho({\bf{y}}) - \delta^{d}({\bf{y}}-{\bf{x}})\}];{\bf{x}}^{'})
 + \Phi([\rho];{\bf{x}}^{'})
 = \pi \mbox{     }m({\bf{x}},{\bf{x}}^{'})
\label{recur}
\end{equation}
 for $ {\bf{x}} \neq {\bf{x}}^{'} $ 
 and  $ m({\bf{x}},{\bf{x}}^{'}) = -m({\bf{x}}^{'},{\bf{x}}) $
 is an odd integer.
 A further constraint on $ \Phi $ emerges
 when we realise that Eq.(~\ref{CURR}) has to be consistent with
 current algebra.
 We know that currents and densities obey the current
 algebra. In other words,
\begin{equation}
[{\bf{J}}({\bf{x}}),\rho({\bf{x}}^{'})] = -i\rho({\bf{x}})
\nabla_{ {\bf{x}} }\delta^{d}({\bf{x}}-{\bf{x}}^{'})
\end{equation}
\begin{equation}
[{\bf{J}}_{i}({\bf{x}}),{\bf{J}}_{j}({\bf{x}}^{'})] = 
 -i{\bf{J}}_{j}({\bf{x}})
\nabla^{i}_{ {\bf{x}} }\delta^{d}({\bf{x}}-{\bf{x}}^{'})
+ i{\bf{J}}_{i}({\bf{x}}^{'})
\nabla^{j}_{ {\bf{x}}^{'} }\delta^{d}({\bf{x}}^{'}-{\bf{x}})
\end{equation}
\begin{equation}
[\rho({\bf{x}}),\rho({\bf{x}}^{'})] = 0
\end{equation}
 In what follows we try and impress upon the reader the need for a systematic
 and general approach by pointing out that reasonable sounding and rather
 general ansatzs fail to obey the constraints just outlined. Thus it seems
 that constraint on $ \Phi $ due to current algebra is almost inconsistent
 with the constraint due to the statistics requirement. 

\subsection{ A Serendipitous Surmise }

Many years ago the author had a conversation with then the student
now Prof. A.H. Castro-Neto where the latter suggested that maybe
the field operator is simply given by,
\begin{equation}
\psi({\bf{x}}) \approx e^{-i \sum_{ {\bf{q}} } e^{ i{\bf{q}}.{\bf{x}} }
X_{ {\bf{q}} } } \sqrt{ \rho^{0} }
\label{SEREN}
\end{equation}
 where by definition $ X_{ {\bf{q}} } =
 i {\bf{q}} \cdot {\bf{j}}(-{\bf{q}})/({\bf{q}}^{2}N^{0}) $
 and $ \rho^{0} = N^{0}/V $.  Later he realised that if we choose
 $ [X_{ {\bf{q}} }, X_{ {\bf{q}}^{'} }] = 0 $ as is in fact
 mandatory \footnote{ Since strictly speaking it is the conjugate to $ \rho $
 as defined by the line integral of the ratio of the current and density
 that enters; these commute amongst themselves. }, then fermion commutation
 rules are not obeyed. However one may take solace in the fact that at least
 one commutation rule does come
 out right namely $ [\psi({\bf{x}}), \rho_{ {\bf{q}} }] = 
e^{ i{\bf{q}}.{\bf{x}} } \psi({\bf{x}}) $. A refinement over this ansatz
 was attempted in our earlier work\cite{Setlur1} by introducing
 an additional phase functional of the density linear in the density and
 this was also inadequate since by now the author knows that $ \Phi $ 
 there when computed was imaginary when it was postulated to be real.
 A compromise was also suggested that involved multiplying and dividing
 by the free propagator and using the exact version in the numerator and
 the bosonized free propagator in the denominator. This trick though
 repugnant to most, gives us an anomalous exponent
 of the Luttinger liquid as we shall see below. However, this anomalous
 exponent is off by a factor of two from the exact one obtained by
 Mattis and Lieb\cite{Mattis}.
 From Eq.(~\ref{SEREN}) we may write,
\[
G({\bf{x}}-{\bf{x}}^{'}) = \left< \psi^{\dagger}({\bf{x}}^{'}) 
\psi({\bf{x}}) \right> 
\approx \left< e^{i \sum_{ {\bf{q}} } 
\left( e^{ i{\bf{q}}.{\bf{x}}^{'} }
 - e^{ i{\bf{q}}.{\bf{x}} } \right)
X_{ {\bf{q}} } } \right>
\mbox{        }
\rho^{0}
\]
\begin{equation}
=  e^{- \sum_{ {\bf{q}} } 
\left( 1 - cos[{\bf{q}}.({\bf{x}}-{\bf{x}}^{'})] \right)
\left< X_{ {\bf{q}} } X_{ -{\bf{q}} } \right> }
\mbox{        }
\rho^{0}
\end{equation}
 Again using the trick outlined in our
 earlier work\cite{Setlur1} we may write,
\begin{equation}
G({\bf{x}}-{\bf{x}}^{'}) = G_{0}({\bf{x}}-{\bf{x}}^{'}) \mbox{      }
e^{- \sum_{ {\bf{q}} } 
\left( 1 - cos[{\bf{q}}.({\bf{x}}-{\bf{x}}^{'})] \right)
[\left< X_{ {\bf{q}} } X_{ -{\bf{q}} } \right>
-\left< X_{ {\bf{q}} } X_{ -{\bf{q}} } \right>_{0}] }
\end{equation}
 Here $ G_{0}({\bf{x}}-{\bf{x}}^{'}) $ is the propagator
 obtained from elementary considerations. In one dimension, we may see
 that $ \left< X_{ {\bf{q}} } X_{ -{\bf{q}} } \right> \approx
 k^{2}_{F} S(q) /(q^{2}N^{0}) $. The structure factor
 $ S_{0}(q) = |q|/(2k_{F}) $ for the interacting case. 
 For the interacting case we have, $ S(q) = (v_{F}/v_{eff}) \mbox{ }S_{0}(q) $.
\[
G(x-x^{'}) = G_{0}(x-x^{'}) \mbox{      }
e^{ -\int^{\infty}_{0} dq \mbox{       } 
\frac{ 1 - cos[q.(x-x^{'})] }{|q|} 
\left( \frac{ v_{F} }{2v_{eff}} - \frac{1}{2} \right) }
\]
\begin{equation}
\sim  G_{0}(x-x^{'}) \mbox{      } \left( \frac{1}{|x-x^{'}|} \right)^{\gamma}
\end{equation}
 where $ \gamma = \frac{ v_{F} }{2v_{eff}} - \frac{1}{2} $.                     This exponent is exactly one half of the exponent 
 obtained by Mattis and Lieb\cite{Mattis}.
 What is even worse is, we have shown in an
 earlier preprint\cite{preprint} 
  that when applied to the X-ray edge problem, we obtain
 the well-known results of Mahan in one dimension but not in higher dimensions.
 Thus it would appear that there is something amiss in the expression
 for the field operator. The present attempt is to finally
 address this difficulty.

\subsection{ No-Go Theorems }

 We make some observations that render seemingly obvious and 
 promising approaches futile.

\vspace{0.2in}

{\bf{The First No-Go Theorem}} : 

\noindent  Here we show that a simple and
 very reasonable ansatz for $ m(x,x^{'}) $
 in one dimension, fails. Set
 $ m(x,x^{'}) = \theta(x-x^{'}) - \theta(x^{'}-x) $. Clearly
 $ m = \pm 1 $ is odd and $ m(x,x^{'}) = -m(x^{'},x) $.
 It may be shown using the method of generating functions that 
 the most general solution to Eq.(~\ref{recur}) is as shown below.
\begin{equation}
\Phi([\rho];x) = \pi \int^{x^{-}}_{-\infty} dy \mbox{        }\rho(y) 
 + C_{0}([\{\rho(y) - \delta(y-x)\}])
 - C_{0}([\rho])
\label{PHIX}
\end{equation}
 Here $ C_{0} $ is arbitrary. Unfortunately, the presence of the first term
 means that current algebra is violated. Using Eq.(~\ref{PHIX}) in
 Eq.(~\ref{CURR}) we have
\begin{equation}
J(x) = -\rho(x)\partial_{x} {\tilde{ \Pi }}(x)
 + \rho(x) \pi \rho(x^{-})  
\label{CURR1D}
\end{equation}
Then the current-current commutator reads as follows.
\[
[J(x),J(x^{'})]  
=  - iJ(x) \partial_{x} \delta(x-x^{'}) 
  + iJ(x^{'})  \partial_{x^{'}} \delta(x^{'}-x)
\]
\begin{equation}
- i\pi \rho(x) \rho(x_{-})
\mbox{  }\partial_{x} \delta(x-x^{'})
+ i\pi \rho(x^{'}) \rho(x^{'}_{-})
\mbox{  }\partial_{x^{'}} \delta(x^{'}-x)
\end{equation}
 The last two terms tell us that something is not right. Thus this attempt
 has come to naught. It is telling us that perhaps $ m $ depends on
 the density as well.

\vspace{0.2in}

{\bf{The Second No-Go Theorem}} : 

\noindent Here we show that a general looking ansatz for $ \Phi $ that
 manifestly obeys current algebra fails to reproduce fermion commutation rules.
\begin{equation}
\Phi([\rho];{\bf{x}}) =
 B_{0}([\{\rho({\bf{y}}) - \delta^{d}({\bf{y}}-{\bf{x}})\}])
 - B_{1}([\rho])
\label{PHICURR}
\end{equation}
 Using the fact that
 $ \nabla B_{0}([\{\rho({\bf{y}}) - \delta^{d}({\bf{y}}-{\bf{x}})\}]) =  [-iB_{0}([\{\rho({\bf{y}}) - \delta^{d}({\bf{y}}-{\bf{x}})\}]),\nabla \Pi({\bf{x}})]  $ (this can be shown easily by Fourier decomposing $ \Phi $ with respect to
 $ \rho $). 
\begin{equation}
{\bf{J}}({\bf{x}}) = -\rho({\bf{x}})\nabla {\tilde{\Pi}}({\bf{x}})
\label{CURRNEW}
\end{equation}
\begin{equation}
{\tilde{\Pi}}({\bf{x}}) = \Pi({\bf{x}}) + [iB_{1}([\rho]),\Pi({\bf{x}})] 
\label{PITILDE}
\end{equation}
 It can be seen that $ {\tilde{\Pi}} $ is also a
 canonical conjugate of $ \rho $.
\begin{equation}
[{\tilde{\Pi}}({\bf{x}}), {\tilde{\Pi}}({\bf{x}}^{'})] = 0
\end{equation}
\begin{equation}
[{\tilde{\Pi}}({\bf{x}}), \rho({\bf{x}}^{'})] = i \mbox{        }
\delta^{d}({\bf{x}}-{\bf{x}}^{'})
\end{equation}
 Thus  Eq.(~\ref{CURRNEW}) obeys current algebra. Unfortunately 
 Eq.(~\ref{PHICURR}) fails to obey  Eq.(~\ref{recur}).
 To see this we merely plug in  Eq.(~\ref{PHICURR}) into Eq.(~\ref{recur})
 and find,
\begin{equation}
F([\rho];{\bf{x}}) - F([\rho];{\bf{x}}^{'})
 =_{?} \pi \mbox{       }(odd \mbox{       }integer)
\label{CONTRA}
\end{equation}
\begin{equation}
 F([\rho];{\bf{x}}) = 
 B_{1}([\{\rho({\bf{y}}) - \delta^{d}({\bf{y}}-{\bf{x}}) \}])
 - B_{0}([\{\rho({\bf{y}}) - \delta^{d}({\bf{y}}-{\bf{x}}) \}])
\end{equation}
 It can be seen that Eq.(~\ref{CONTRA}) is a contradiciton. 
 If $ F([\rho];{\bf{x}})/\pi $ is odd then $ F([\rho];{\bf{x}}^{'})/\pi $
 has to be even for {\it{all}} $ {\bf{x}}^{'} \neq {\bf{x}} $.
 In particular, $  F([\rho];{\bf{x}}_{1}^{'})/\pi $ and
 $ F([\rho];{\bf{x}}_{2}^{'})/\pi $ have to be both even.
 But now their difference can no longer be odd, thus violating the recursion
 relation. While this is somewhat disappointing, it is not surprising
 since Eq.(~\ref{CURRNEW}) suggests that the velocity is irrotational, namely,
 $ {\bf{v}} = -\nabla {\tilde{\Pi}} $. This is clearly not general enough.
 We would like to be able to write
 $ {\bf{v}} = -\nabla {\tilde{\Pi}} + {\bf{v}}^{r}$. In what follows we shall
 do precisely this.

\vspace{0.2in}

\noindent We set $ {\bf{J}} = \rho \mbox{    }{\bf{v}} $
   and ask what properties should $ {\bf{v}} $ possess ?
   This implies the following commutation rules on the velocity operator.
\begin{equation}
[{\bf{v}}({\bf{x}}),\rho({\bf{x}}^{'})] = -i
\nabla_{ {\bf{x}} }\delta^{d}({\bf{x}}-{\bf{x}}^{'})
\end{equation}
and,
\[
 -i\rho({\bf{x}})\delta^{d}({\bf{x}}-{\bf{x}}^{'})
 \nabla^{i}_{ {\bf{x}} }{\bf{v}}_{j}({\bf{x}})
+i \rho({\bf{x}}^{'})
\delta^{d}({\bf{x}}-{\bf{x}}^{'})  \nabla^{j}_{ {\bf{x}}^{'} }
 {\bf{v}}_{i}({\bf{x}}^{'})  
\]
\begin{equation}
+ \rho({\bf{x}}) \rho({\bf{x}}^{'})
[{\bf{v}}_{i}({\bf{x}}), {\bf{v}}_{j}({\bf{x}}^{'})] 
 = 0
\end{equation}
 From Eq.(~\ref{CURR}) it is clear that we may expect to be able
 to write the velocity as a sum of two parts, one irrotational
 and the other depending only on the density which has a nonvanishing curl. 
 Thus,
\begin{equation}
{\bf{v}} = {\bf{v}}^{s} + {\bf{v}}^{r}
\end{equation}
 Here $ {\bf{v}}^{s} = - \nabla {\tilde{\Pi}}^{s} $
 for some $ {\tilde{\Pi}}_{s} $. Without
 loss of generality we may choose
 $ [{\bf{v}}_{i}^{s}({\bf{x}}), {\bf{v}}_{j}^{s}({\bf{x}}^{'} )] = 0 $.
 In other words we include only those parts of the velocity into the
 irrotational part that makes it commute amongst itself. 
 Of course, $  [{\bf{v}}_{i}^{r}({\bf{x}}),  {\bf{v}}_{j}^{r}({\bf{x}}^{'})] = 0 $. Crucially however, $  [{\bf{v}}_{i}^{r}({\bf{x}}), {\bf{v}}_{j}^{s}({\bf{x}}^{'})] \neq 0 $. Indeed,
\begin{equation}
 [{\bf{v}}_{i}^{r}({\bf{x}}), {\bf{v}}_{j}^{s}({\bf{x}}^{'})]
 = -\nabla^{j}_{ {\bf{x}}^{'} }
[{\bf{v}}_{i}^{r}({\bf{x}}), i \frac{ \delta }{ \delta \rho({\bf{x}}^{'}) }]
\end{equation}
\begin{equation}
[{\bf{v}}^{s}({\bf{x}}),\rho({\bf{x}}^{'})] = -i
\nabla_{ {\bf{x}} }\delta^{d}({\bf{x}}-{\bf{x}}^{'})
\end{equation}
and,
\[
 - i\rho({\bf{x}})\delta^{d}({\bf{x}}-{\bf{x}}^{'})
 \nabla^{i}_{ {\bf{x}} }{\bf{v}}^{r}_{j}({\bf{x}})
+ i \rho({\bf{x}}^{'})
\delta^{d}({\bf{x}}-{\bf{x}}^{'})  \nabla^{j}_{ {\bf{x}}^{'} }
 {\bf{v}}^{r}_{i}({\bf{x}}^{'})  
\]
\begin{equation}
- \rho({\bf{x}}) \rho({\bf{x}}^{'})\nabla^{i}_{ {\bf{x}} }
[i \frac{ \delta }{ \delta \rho({\bf{x}}) }
, {\bf{v}}^{r}_{j}({\bf{x}}^{'})] 
- \rho({\bf{x}}) \rho({\bf{x}}^{'})
\nabla^{j}_{ {\bf{x}}^{'} }
[{\bf{v}}^{r}_{i}({\bf{x}}),
 i \frac{ \delta }{ \delta \rho({\bf{x}}^{'}) }] = 0
\end{equation}
This means that,
\begin{equation}
 - i\frac{ \delta^{d}({\bf{x}}-{\bf{x}}^{'}) }{\rho({\bf{x}})}
 \nabla^{i}_{ {\bf{x}} }{\bf{v}}^{r}_{j}({\bf{x}})
- \nabla^{i}_{ {\bf{x}} }
[i \frac{ \delta }{ \delta \rho({\bf{x}}) }
, {\bf{v}}^{r}_{j}({\bf{x}}^{'})] = f_{ij}([\rho];{\bf{x}};{\bf{x}}^{'})
\label{CONSTR}
\end{equation}
for some $ f_{ij}([\rho];{\bf{x}};{\bf{x}}^{'}) = f_{ji}([\rho];{\bf{x}}^{'};{\bf{x}}) $.
 Therefore so long as Eq.(~\ref{CONSTR}) is obeyed then current algebra is
 also respected. In light of the two no-go theorems let us
 set $ \Phi $ to be as shown below.
\begin{equation}
  \Phi([\rho];{\bf{x}}) = \Phi_{irr}([\rho];{\bf{x}})
 + B_{0}([\{\rho({\bf{y}}) - \delta^{d}({\bf{y}}-{\bf{x}})\}])
 - B_{0}([\rho])
\end{equation}
 Here $ \Phi_{irr} $ stands for some particular solution and
 $ \Phi $ would be a general solution. If $ \Phi_{irr} $ obeys current
 algebra and the recursion relation so too does $ \Phi $.
 From Eq.(~\ref{CURR}) we may read off the following formula for
 the velocity operator $ {\bf{v}} = {\bf{v}}^{s} + {\bf{v}}^{r} $.
 Where,
\begin{equation}
{\bf{v}}^{r}({\bf{x}}) = 
 \nabla \Phi_{irr}([\rho];{\bf{x}}) -
  [\Phi_{irr}([\rho];{\bf{x}}),\nabla_{ {\bf{x}} }
\mbox{      } \frac{ \delta }{ \delta \rho({\bf{x}}) }] 
\end{equation}
 and $ {\bf{v}}^{s} = -\nabla {\tilde{ \Pi }}^{s} $ 
 where $ {\tilde{\Pi}}^{s} $ is given by Eq.(~\ref{PITILDE}). 
 We are looking for a singular solution to $ \Phi $ that also obeys the
 statistics requirement. Anticipating some simplifications,
 let us reparametrise $ \Phi $ as follows.
\begin{equation}
\Phi_{irr}([\rho];{\bf{x}}) = \pi \mbox{        }
L([\{\rho({\bf{y}}) - \delta^{d}({\bf{y}}-{\bf{x}})\}];{\bf{x}})
 - \pi \mbox{      }L([\rho];{\bf{x}})
\label{PHIIRR}
\end{equation}
 In fact it can be shown that only a form such as in
 Eq.(~\ref{PHIIRR}) is consistent with the statistics requirement
 in Eq.(~\ref{recur}). 
 The statistics requirement may now be written in exponential form
 as follows.
\[
e^{- i \pi L([\{\rho({\bf{y}}) - \delta^{d}({\bf{y}}-{\bf{x}}^{'})\}];{\bf{x}}) }
e^{ i \pi L([\rho];{\bf{x}}) }
e^{i \pi L([\{\rho({\bf{y}}) - \delta^{d}({\bf{y}}-{\bf{x}})\}];{\bf{x}}^{'}) }
e^{- i \pi L([\rho];{\bf{x}}^{'}) }
\]
\[
 = - \mbox{        }
e^{-i \pi L([\{\rho({\bf{y}}) - \delta^{d}({\bf{y}}-{\bf{x}}^{'})
- \delta^{d}({\bf{y}}-{\bf{x}}) \}];{\bf{x}}) }
e^{ i \pi L([\{\rho({\bf{y}}) - \delta^{d}({\bf{y}}-{\bf{x}})\}];{\bf{x}}) }
\]
\begin{equation}
\times \mbox{        }
e^{i \pi L([\{\rho({\bf{y}}) 
 -  \delta^{d}({\bf{y}}-{\bf{x}}^{'})
- \delta^{d}({\bf{y}}-{\bf{x}})\}];{\bf{x}}^{'}) }
e^{- i \pi L([\{\rho({\bf{y}}) - \delta^{d}({\bf{y}}-{\bf{x}}^{'})\}];{\bf{x}}^{'}) }
\label{STATREQ}
\end{equation}
 The requirement in Eq.(~\ref{STATREQ}) is unfortunately
 strictly speaking, inconsistent with the constraint from current algebra. 
 To see this, we note that in one dimension,
 the constraint from current algebra reads as follows.
\begin{equation}
[-i \partial_{x} \frac{ \delta }{ \delta \rho(x) }, v^{r}(x^{'})]
 = [-i \partial_{x^{'}} \frac{ \delta }{ \delta \rho(x^{'}) }, v^{r}(x)]
\end{equation}
This can be satisfied only if,
\begin{equation}
 v^{r}(x) =  -\pi \mbox{      }
[\partial_{x}  \frac{ \delta }{ \delta \rho(x) }, 
L([\rho])]
\label{eqvr}
\end{equation}
 for some $ L $ (see appendix A).
 We see that this means that the $ L $ of Eq.(~\ref{PHIIRR})
 has to be now independent of $ {\bf{x}} $. 
 This means unfortunately that  Eq.(~\ref{STATREQ}) is violated. 
 However, one can take the point of view that we would like the current
 algebra to be `almost obeyed', in other words we allow $ L $ 
 to depend extremely weakly on $ {\bf{x}} $. This will also open up the
 possibility that  Eq.(~\ref{STATREQ}) may now be satisfied.

\section{ Evaluating the Phase Functional }

 Here we would like to compute the phase functional that obeys the 
 constraints just described. Computing the phase functional
 systematically seems particularly difficult.
 Intuition suggests that the phase functional
 ought to be related to the properties of the free theory.
 Specifically, we should be able
 expand it in powers of the density fluctuations and somehow relate the
 coefficients to the density-density, current-density
 and current-current correlation functions. From the preceeding
 sections, it appears that we may write,
\begin{equation}
\psi({\bf{x}}) = e^{i \Lambda([\rho];{\bf{x}}) }
 e^{ -i \Pi({\bf{x}}) } \sqrt{ \rho({\bf{x}}) }
\label{EQNPSIN}
\end{equation}  
 and $ \Lambda $ is some functional that depends weakly on
 $ {\bf{x}} $. In fact it has to be strictly
 independent of $ {\bf{x}} $ in order for
 current algebra to be obeyed. 
 However, the weak dependence is needed to recover
 Fermi statistics.  Indeed if we choose $ \Lambda $ to be independent of
 $ {\bf{x}} $ and then evaluate the equal-time version 
 of the  propagator of the Luttinger model, we find an
 anomalous exponent equal to one-half of the exact one derived by
 Mattis and Lieb.
 In other words, an $ \Lambda $ independent of $ {\bf{x}} $
 is more or less equivalent to the serendipitous
 surmise. Thus we have to do better.
 In general $ \Pi $ may be written as
 $ \Pi({\bf{x}}) = X_{0} + {\tilde{ \Pi }}({\bf{x}}) $. Here
 $ X_{0} $ is conjugate to the total number of particles and 
 $ {\tilde{ \Pi }} $ is strictly selfadjoint since it is related to
 currents and densities. We shall assume that $ X_{0} $ is also
 self-adjoint but may not be expressed in terms of Fermi bilinears.
 It may be shown\cite{Setlur1} that $ X_{0} $ has an expression shown below.
\begin{equation}
X_{0} = \frac{i}{2N^{0}}\sum_{ {\bf{k}} } 
n_{F}({\bf{k}}) \mbox{       }Ln( \mbox{  }c_{ {\bf{k}} } \mbox{  })
 - \frac{i}{2N^{0}}\sum_{ {\bf{k}} } 
n_{F}({\bf{k}}) \mbox{       }Ln( \mbox{  }c^{\dagger}_{ {\bf{k}} } \mbox{  })
\label{EQNX0}
\end{equation}
 It seems that the lagrangian formulation is appropriate here.
 The hamiltonian formulation has not worked out for unknown reasons.
 In the lagrangian formulation, all the variables are c-numbers.
 In the usual path integral formulation, the Fermi fields are 
 complex Grassmann numbers and not ordinary complex numbers.
 As pointed out earlier,
 we take the point of view that a complex Grassmann number field 
 may be `simulated' by a complex ordinary number field provided one 
 introduces an appropriate phase functional $ \Lambda $.
 The value of $ \Lambda $
 is tuned so as to reproduce the correlation functions of the free theory.
 The anticommutating nature of the Grassmann number is not 
 present in the polar decomposition in Eq.(~\ref{EQNPSIN}) but is
 recovered at the level of the propagator.
 To see this we note that in the path integral formulation the fields
 obey periodic (KMS)  boundary conditions\cite{Kadanoff} namely,
\begin{equation}
\psi({\bf{x}},t-i\beta) = - e^{ \beta \mu } \mbox{        }
\psi({\bf{x}},t)
\label{KMS}
\end{equation}
 Using Eq.(~\ref{KMS}) in Eq.(~\ref{EQNX0}) we find,
\begin{equation}
X_{0}(t-i\beta) = X_{0}(t) + i \mbox{       }
Log \left( - e^{ \beta \mu }\right)
 = X_{0}(t) - \pi + i \beta \mu 
\end{equation}
 That is, the one-particle propagators obey Fermi commutation rules mainly
 due to the fermionic boundary conditions obeyed by the global Klein factor.
 The current operator may be written as follows.
\begin{equation}
{\bf{J}}({\bf{x}}) = - \rho({\bf{x}}) \nabla \Pi({\bf{x}}) +
 \rho({\bf{x}})  {\bf{V}}([\{\rho_{ {\bf{q}} } - e^{ i {\bf{q}}.{\bf{x}} }\}];{\bf{x}})
\end{equation}
where,
\begin{equation}
\nabla \Lambda([\rho];{\bf{x}}) = 
{\bf{V}}([\rho];{\bf{x}}) 
\end{equation}
 First we note that in order for current algebra to be obeyed,
 $ \Lambda([\rho];{\bf{x}}) $ has to be independent of $ {\bf{x}} $.
 In other words $ {\bf{V}}([\rho];{\bf{x}})  \equiv 0 $. Unfortunately
 this means that Fermi statistics is no longer obeyed. 
 However, we can take the point of view that 
 $ \Lambda $ depends very weakly on $ {\bf{x}} $ and thus we may ignore
 the derivative of $ \Lambda $ since it involves taking the difference between
 neighboring points. On the other hand, for statistics we are
 usually interested in the opposite extreme namely the asymptotic regime
 $ |{\bf{x}}-{\bf{x}}^{'}| \rightarrow \infty $. In other words, we may
 no longer ignore
 $ \Lambda([\rho];{\bf{x}}) - \Lambda([\rho];{\bf{x}}^{'}) $
 for large separations. However ignoring $ {\bf{V}} $ also seems
 to be a bad idea. For example if we do and use the Dashen-Sharp formula
 for the kinetic energy we get a hamiltonian that describes bosons
 rather than fermions. 
\[
K = \int \frac{ d^{d}x }{2m}
\left( \frac{ {\bf{J}}^2 }{\rho} + \frac{ (\nabla \rho)^2 }{4 \rho }
 \right)
\]
\begin{equation}
\int \frac{ d^{d}x }{2m}
\left( \rho (\nabla \Pi)^2  + \frac{ (\nabla \rho)^2 }{4 \rho }
 \right)
\approx \sum_{ {\bf{q}} } N \mbox{      }
 \epsilon_{ {\bf{q}} } \mbox{    }X_{ {\bf{q}} }X_{ -{\bf{q}} }
 + \sum_{ {\bf{q}} \neq 0 } \frac{ \epsilon_{ {\bf{q}} } }{4N}
\rho_{ {\bf{q}} } \rho_{ -{\bf{q}} }
\end{equation}
Thus it appears that we have to retain $ {\bf{V}} $ as well.
We would like to write down the action in the lagrangian formulation.
For this we first write,
\begin{equation}
H =  \int \frac{ d^{d}x }{2m}
\left( \rho (\nabla \Pi - {\bf{V}}^{'})^2  + \frac{ (\nabla \rho)^2 }{4 \rho }
 \right)
\label{HQU}
\end{equation}
 where,
\begin{equation}
{\bf{V}}^{'}([\rho];{\bf{x}}) = 
 {\bf{V}}([\{\rho_{ {\bf{q}} } - e^{ i {\bf{q}}.{\bf{x}} }\}];{\bf{x}})
\end{equation}
 Till now the discussion has been at the quantum level. 
 Now we would like to recast this theory in the path integral language. For
 this we have to compute the classical action in terms of the 
 collective coordinates $ \Pi, \rho $. This means we have to 
 `de-quantize' the quantum hamiltonian in Eq.(~\ref{HQU}). 
 Happily, while there are many inequivalent ways of quantizing a given
 classical hamiltonian there is likely a unique way of de-quantizing 
 a quantum hamiltonian.  Thus we may make the following observations.
 The Hamilton equations for the canonical variables are,
\begin{equation}
\partial_{t} \Pi({\bf{x}},t) = 
-\frac{ \delta H(\rho,\Pi) }{ \delta \rho({\bf{x}},t) }
\end{equation}
\begin{equation}
\partial_{t} \rho({\bf{x}},t) = 
\frac{ \delta H(\rho,\Pi) }{ \delta \Pi({\bf{x}},t) }
\end{equation}


\begin{equation}
\partial_{t} \rho({\bf{x}},t) = 
- \frac{1}{2m}
\nabla_{ {\bf{x}} }
\left( 2 \rho({\bf{x}}) ( \nabla_{ {\bf{x}} }
 \Pi({\bf{x}}) - {\bf{V}}^{'}([\rho];{\bf{x}}) )  \right)
\end{equation}

\begin{equation}
L(\rho,\partial_{t} \rho) = \int d^{d}x \mbox{       }\Pi({\bf{x}},t) 
 \partial_{t} \rho({\bf{x}},t) 
 - H(\rho,\Pi)
\end{equation}

\[
L(\rho,\partial_{t} \rho) = -\int d^{d}x \mbox{       }\Pi({\bf{x}},t) 
 \frac{1}{2m} \nabla_{ {\bf{x}} }
\left( 2 \rho({\bf{x}}) ( \nabla_{ {\bf{x}} }
 \Pi({\bf{x}}) - {\bf{V}}^{'}([\rho];{\bf{x}}) )  \right)
 -  \int \frac{ d^{d}x }{2m}
\left( \rho (\nabla \Pi - {\bf{V}}^{'})^2  + \frac{ (\nabla \rho)^2 }{4 \rho }
 \right)
\]
\[
 = \frac{1}{m} 
 \int d^{d}x \mbox{       } \nabla_{ {\bf{x}} } \Pi({\bf{x}},t) 
\left( \rho({\bf{x}}) ( \nabla_{ {\bf{x}} }
 \Pi({\bf{x}}) - {\bf{V}}^{'}([\rho];{\bf{x}}) )  \right)
 -  \int \frac{ d^{d}x }{2m}
\left( \rho (\nabla \Pi - {\bf{V}}^{'})^2  + \frac{ (\nabla \rho)^2 }{4 \rho }
 \right)
\]
\begin{equation}
 =  \frac{1}{m} 
 \int d^{d}x \mbox{       }   \rho({\bf{x}}) \mbox{        }
{\bf{V}}^{'}([\rho];{\bf{x}})\mbox{        }
 ( \nabla_{ {\bf{x}} }
 \Pi({\bf{x}}) - {\bf{V}}^{'}([\rho];{\bf{x}}) ) 
 + \int \frac{ d^{d}x }{2m}
\left( \rho (\nabla \Pi - {\bf{V}}^{'})^2 - \frac{ (\nabla \rho)^2 }{4 \rho }
 \right)
\end{equation}

\section{ One Dimension }

In one dimension we have,
\begin{equation}
\partial_{t} \rho(x,t) = 
- \frac{1}{m}
\partial_{x}
\left( \rho(x) ( \partial_{x}
 \Pi(x) - V^{'}([\rho];x) )  \right)
\end{equation}
\begin{equation}
( \partial_{x}
 \Pi(x,t) - V^{'}([\rho];x) )
 = - \frac{m}{ \rho(x,t) }  \int^{x}_{ -\infty } dx^{'} \mbox{      }
\partial_{t} \rho(x^{'},t)
\end{equation}
The Lagrangian is given by,
\[
L(\rho,\partial_{t} \rho) =
- \int_{-\infty}^{\infty}  dx  \mbox{        }
V^{'}([\rho_{t}];x)\mbox{        }
  \int^{x}_{ -\infty } dx^{'} \mbox{      }
\partial_{t} \rho(x^{'},t)
\]
\begin{equation}
 + \int_{ -\infty }^{ \infty } \frac{ dx }{2m}
\left( \frac{ m^2 }{ \rho(x,t) }
 \left( \int^{x}_{ -\infty }dx^{'} \partial_{t} \rho(x^{'},t)
 \right)^2  - \frac{ ( \partial_{x} \rho(x,t) )^2 }{4 \rho(x,t) }
 \right)
\end{equation}
The current operator is then given simply by,
\begin{equation}
J(x,t) = m \int^{x}_{ - \infty } dx^{'} \partial_{t} \rho(x^{'},t) 
\end{equation}
 So far the discussion has been exact. Now we would like to make some
 approximations. In particular we assume that we are in the degenerate regime
 so that $ \rho(x,t) = \rho_{0} + {\tilde{\rho}}(x,t) $ 
 and $  {\tilde{\rho}}(x,t) \ll \rho_{0} $ and $ \rho_{0} = N^{0}/L $ 
 the number of electrons per unit length.
\[
L(\rho,\partial_{t} \rho) =
- \int_{-\infty}^{\infty}  dx  \mbox{        }
V^{'}([\rho_{0} + {\tilde{\rho}}_{t}];x)\mbox{        }
  \int^{x}_{ -\infty } dx^{'} \mbox{      }
\partial_{t} {\tilde{\rho}}(x^{'},t)
\]
\begin{equation}
 + \int_{ -\infty }^{ \infty } \frac{ dx }{2m}
\left( \frac{ m^2 }{ \rho_{0} }
 \left( \int^{x}_{ -\infty }dx^{'} \partial_{t} {\tilde{\rho}}(x^{'},t)
 \right)^2  - \frac{ ( \partial_{x} {\tilde{\rho}}(x,t) )^2 }{4 \rho_{0} }
 \right)
\end{equation}

\begin{equation}
{\tilde{\rho}}(x,t) = \frac{1}{L} \sum_{ q \neq 0 } \rho_{q}(t)
 \mbox{       }e^{-iqx} 
\end{equation}
\[
L(\rho,\partial_{t} \rho) =
- \int_{-\infty}^{\infty}  dx  \mbox{        }
V^{'}([\rho_{0} + {\tilde{\rho}}_{t}];x)\mbox{        }
  \int^{x}_{ -\infty } dx^{'} \mbox{      }
\partial_{t} {\tilde{\rho}}(x^{'},t)
\]
\begin{equation}
 + \int_{ -\infty }^{ \infty } \frac{ dx }{2m}
\frac{ m^2 }{ \rho_{0} }
 \left( \int^{x}_{ -\infty }dx^{'} \partial_{t} {\tilde{\rho}}(x^{'},t)
 \right)^2  
- \sum_{ q \neq 0 } \frac{ \epsilon_{q} }{4 N^{0} } \rho_{q}(t) \rho_{-q}(t)
\end{equation}

\[
 \int^{x}_{ -\infty }dx^{'} \partial_{t} {\tilde{\rho}}(x^{'},t)
 =  \frac{1}{L} \sum_{ q \neq 0 }  \partial_{t} \rho_{q}(t)
 \mbox{       }\frac{ e^{-iqx} }{ -i q }  
\]

\[
L(\rho,\partial_{t} \rho) =
 \sum_{q \neq 0} v^{'}([\rho_{0} + {\tilde{\rho}}_{t}];q)\mbox{        }
  \frac{ \partial_{t} \rho_{q}(t) }{ i q }  
\]
\begin{equation}
 +  \sum_{ q \neq 0 }  \frac{ \partial_{t} \rho_{q}(t)
 \partial_{t} \rho_{-q}(t) }{ 4 N^{0} \epsilon_{q} }  
- \sum_{ q \neq 0 } \frac{ \epsilon_{q} }{4 N^{0} } \rho_{q}(t) \rho_{-q}(t)
\end{equation}

\begin{equation}
j_{q}(t) = \frac{ i m }{q} \mbox{      }
\partial_{t} \rho_{q}(t) 
\end{equation}

\begin{equation}
v^{'}([\rho_{0} + {\tilde{\rho}}_{t}];q) = 
\int_{-\infty}^{\infty}  \frac{ dx }{L} \mbox{        }
V^{'}([\rho_{0} + {\tilde{\rho}}_{t}];x)\mbox{        }
 e^{-iqx}
\end{equation}
 We would now like to determine $ v^{'} $ and hence $ \Lambda $ 
 by forcing this action to reproduce
 the correct current-current, current-density
 and density-density correlation functions of the noninteracting Fermi theory. 
 Unfortunately since the time dependence of $ v^{'} $ is through $ \rho $
 alone this is not possible as we have found after repeated attempts
 by choosing various $ v^{'} $, linear, quadratic and so on in $ \rho $.
 Thus we now have to form another reinterpretation. We have the luxury
 of reinterpreting these formulas anyway we wish so long as the properties
 of the free theory are properly recovered. Thus we shall transfer the
 time dependence on to the function $ \Lambda $ itself thereby making the  
 action depend on the history of the density of the system.

\section{ Lagrangian Formulation }

 The discussion in the above sections shows that for fermions, starting from
 a Hamiltonian formulation and then moving to the Lagrangian formulation
 does not work out.  Thus we would like to start with a Lagrangian formulation
 at the outset. For this we postulate that the field operator is given by,
\begin{equation}
\psi({\bf{x}}t) = e^{ i \Lambda([\rho];{\bf{x}}t) } 
\mbox{       }
 e^{-i \Pi({\bf{x}}t) }
 \sqrt{ \rho({\bf{x}}t) }
\label{FOP}
\end{equation}
 The notation $ \Lambda([\rho];{\bf{x}}t) $ is meant to imply that
 the phase functional potentially 
 depends on the history of the density configurations
 of the system. It does not imply that the phase functional is explicitly
 time dependent. 
 Also $ \Lambda $ depends extremely weakly on $ {\bf{x}} $ thus
 we may ignore $ \nabla \Lambda \approx 0 $. However as pointed out
 before, we may not ignore $ \Lambda([\rho];{\bf{x}}t) 
-  \Lambda([\rho];{\bf{x}}^{'}t)  $
  for $ |{\bf{x}} - {\bf{x}}^{'}| \rightarrow \infty $.
 All the variables are c-nmbers in the Lagrangian formulation.
 Even the field variable in Eq.(~\ref{FOP}) is a c-number.
 The anticommuting nature of the field variable is captured in the 
 global Klein factor as already pointed out earlier.
 The current is now given in terms of the canonical variables
 as follows. Since we may ignore the gradient of $ \Lambda $ we have
\begin{equation}
{\bf{J}}({\bf{x}}t) = - \rho({\bf{x}}t) \nabla \Pi({\bf{x}}t)
\end{equation}
 The action may now be recast as follows.
\begin{equation}
S_{free} = \int^{-i\beta}_{0} dt  \int d^{d}x \mbox{        }
\left[ \rho \partial_{t} \Pi - \rho \partial_{t} \Lambda
- \rho ( \nabla \Pi )^2 - \frac{ (\nabla \rho)^2 }{ 4 \rho } 
 \right]  
\label{FREEFERMI}
\end{equation}
 After this reinterpretation we would now like the action to be purely
 quadratic in the canonical variables and yet be able to reproduce
 the correct fermion correlation functions. Thus we make the following 
 assumption. In the high density limit we may ignore the
 density fluctuations whenever appropriate and this is equivalent to
 working in the asymptotic regime.
\begin{equation}
\rho({\bf{x}}t) = \frac{1}{V} \sum_{ {\bf{q}} n }
 e^{ -i {\bf{q.x}} } \rho_{ {\bf{q}} n } e^{ -z_{n} t }
\end{equation}
\begin{equation}
\Pi({\bf{x}}t) =  \sum_{ {\bf{q}} n }
 e^{ i {\bf{q.x}} } X_{ {\bf{q}} n } e^{ z_{n} t }
\end{equation}
\begin{equation}
\Lambda([\rho];{\bf{x}}t) =  \sum_{ {\bf{q}} \neq 0 n }
 e^{ i {\bf{q.x}} } \mbox{        }
\lambda([\rho];{\bf{q}}n) \mbox{        }
 e^{ z_{n} t }
\end{equation}
 The Klein factor which is the position in dependent part of $ \Pi $ 
 ensures that the fermion KMS boundary conditions are
 obeyed by the field variable since we still have $ z_{n} = 2 \pi n/ \beta $.
 Here $ \rho $ is a real function hence
 $ [ \rho({\bf{x}},t) ]^{*} = \rho({\bf{x}},t^{*}) =  \rho({\bf{x}},-t) $
 since $ t \in [0, -i\beta] $. This means
 $ \rho^{*}_{ {\bf{q}}, n } = \rho_{ -{\bf{q}}, n } $. 
 For a similar reason $ X^{*}_{ {\bf{q}}, n } = X_{ -{\bf{q}}, n } $.
 Note however that $ \Pi $ is not self adjoint due to the
 first ($ {\bf{q}} = 0 $) term, namely the non-self adjoint Klein factor. The
 position dependent part of $ \Pi $ is self adjoint. Also we have
 $ [ \lambda([\rho];{\bf{q}},n) ]^{*} =  \lambda([\rho];-{\bf{q}},n) $.

 The action in Eq.(~\ref{FREEFERMI}) would be identical to the bosonic 
 action were it not for the term
 $ - \int \mbox{       }\rho \partial_{t} \Lambda $. Thus this term
 is crucial and leads to the fermionic correlation functions for the
 current-current, current-density and density-density correlations
 since the current is still given simply as in the bosonic case.
 The action may now be written similar to the bosonic case,
\[
S_{free} = \sum_{ {\bf{q}} n } (-i \beta z_{n}) \rho_{ {\bf{q}} n } 
X_{ {\bf{q}} n } 
 + i \beta N^{0} \sum_{ {\bf{q}}n } {\bf{q}}^2 
 X_{ {\bf{q}} n }  X_{ -{\bf{q}}, -n } 
\]
\begin{equation}
 + \frac{ i \beta }{ 4 N^{0} }
\sum_{ {\bf{q}}n } {\bf{q}}^2  
 \rho_{ {\bf{q}}, n }
 \rho_{ -{\bf{q}}, -n }
 + i \beta  \sum_{ {\bf{q}} \neq 0 n }
z_{n}  \mbox{        }
\rho_{ {\bf{q}} n }  \lambda([\rho];{\bf{q}}n) \mbox{        }
\end{equation}
 where $ \rho^{0} = N^{0}/V $ is the mean density of fermions.
 We have included the leading anharmonic corrections since
 this is going to be important in future works where we may choose to go 
 beyond RPA. We choose,
\begin{equation}
\lambda([\rho];{\bf{q}}n) = C({\bf{q}}n) \mbox{        }
\rho_{ -{\bf{q}}, -n } 
 + \delta_{ n, 0 } \mbox{         }
D([\rho];{\bf{q}})
\end{equation}
 We would like to compute the unknowns $ C $ and $ D $ in the above 
 ansatz by forcing this action to recover the right dynamical current-current
 density-density and current-density correlation functions of the free
 Fermi theory. 
\[
S_{free} = \sum_{ {\bf{q}} n } (-i \beta z_{n}) \rho_{ {\bf{q}} n } 
X_{ {\bf{q}} n } 
 + i \beta N^{0} \sum_{ {\bf{q}}n } {\bf{q}}^2 
 X_{ {\bf{q}} n }  X_{ -{\bf{q}}, -n } 
\]
\begin{equation}
 + \frac{ i \beta }{ 4 N^{0} }
\sum_{ {\bf{q}}n } {\bf{q}}^2  
 \rho_{ {\bf{q}}, n }
 \rho_{ -{\bf{q}}, -n }
 + i \beta  \sum_{ {\bf{q}} \neq 0 n }
z_{n}  \mbox{        }
  C({\bf{q}}n) \mbox{        }
\rho_{ {\bf{q}} n }
\rho_{ -{\bf{q}}, -n } 
\end{equation}
We may now integrate out $ X $ to arrive at an effective action,
\[
S_{eff} = i \beta \sum_{ {\bf{q}} n }   
\left( \frac{ z^2_{n} }{ 4 N^{0} {\bf{q}}^2 } \right) \mbox{       }
\rho_{ {\bf{q}} n }
\rho_{ -{\bf{q}}, -n } 
\]
\begin{equation}
 + \frac{ i \beta }{ 4 N^{0} }
\sum_{ {\bf{q}}n } {\bf{q}}^2  
 \rho_{ {\bf{q}}, n }
 \rho_{ -{\bf{q}}, -n }
 + i \beta  \sum_{ {\bf{q}} \neq 0 n }
z_{n}  \mbox{        }
  C({\bf{q}}n) \mbox{        }
\rho_{ {\bf{q}} n }
\rho_{ -{\bf{q}}, -n } 
\end{equation}
Thus we have,
\begin{equation}
2 \left< \rho_{ {\bf{q}}, n } \rho_{ -{\bf{q}}, -n } \right>_{0} 
 = \frac{1}{ \frac{ \beta z^2_{n} }{ 4 N^{0} {\bf{q}}^2 } 
+  \frac{ \beta {\bf{q}}^2 }{ 4 N^{0} } 
 + \beta z_{n} \mbox{      } C({\bf{q}}n) }  
\end{equation}
 From this, we may read off a formula for $ C $ in terms of the density-density
 correlation function we already know how to compute.
\begin{equation}
 \beta z_{n} \mbox{      } C({\bf{q}}n) 
 = \frac{1}{ 2 \left< \rho_{ {\bf{q}}, n } \rho_{ -{\bf{q}}, -n } \right>_{0} }
- \frac{ \beta z^2_{n} }{ 4 N^{0} {\bf{q}}^2 } 
-  \frac{ \beta {\bf{q}}^2 }{ 4 N^{0} } 
\end{equation}
In one dimension we may write,
\begin{equation}
C(q,n) = \frac{ v^2_{F} }{ 4 N^{0} z_{n} }
\end{equation}
where $ v_{F} $ is the Fermi velocity.
Since we taken care to ensure that current algebra is obeyed we may expect
to find that the current-current, current-density and current-current
correlation functions are also properly recovered. 
 The above action does not give everything right. 
 For example, the three body correlations, 
\mbox{{ $ \left< \rho_{ {\bf{q}}, n } 
\rho_{ -{\bf{q}} + {\bf{q}}^{'}, - n+n^{'} } 
\rho_{ -{\bf{q}}^{'}, -n^{'} } \right> $ }} are zero from the above action
 but clearly they are nonzero in the full free Fermi theory. This means 
 we have to go beyond the quadratic action to capture three and higher body
 correlations. This is not important in the RPA sense and shall ignore this.

 Finally, we have to make sure we are able to compute the full propagator.
 Unfortunately,
 a straightforward application of the formula for $ \lambda $ does not
 yield the right propagator. Thus we have to redefine the field as follows.
 We may decompose the field variable into fast and slow modes.
 It is the slow modes that contribute to the action.
 But the fast modes are needed in the one-particle propagator.
 Thus we write, $ \psi = \Psi_{fast} \mbox{      }\psi_{slow} $.
 We postulate that the fast modes are not affected by interactions, since
 the action only involves hydrodynamic degrees of freedom, in other
 words the slow modes. In other words, 
\begin{equation}
\psi({\bf{x}}t) =  \Psi_{fast}({\bf{x}}t) \mbox{      }
\mbox{        }
e^{ -i \sum_{ {\bf{q}} \neq 0, n } e^{ i {\bf{q}}.{\bf{x}} } e^{ z_{n} t } 
\left( 
X_{ {\bf{q}} n } - G({\bf{q}} n) \mbox{       }\rho_{ -{\bf{q}}, -n } \right) }
\end{equation}
 Here $ G({\bf{q}}n) $ has to be recomputed since we are unable
 to make contact with the free propagator otherwise.
 Perhaps this suggests that the real theory will involve some very complicated
 forms of $ \Lambda $.
 The main aim of this preprint is to convince the reader of the urgency
 of somewhow making this scheme work for then we will be able to
 compute the observable properties of a theory of neutral matter where the
 only adjustable parameters are the relative concentrations of the 
 various elements that make up the material.
 Now we specialise to one dimension where the analysis is somewhat simpler.
 In furture versions of this preprint, we address the 3d case and apply 
 it to the electron-phonon problem briefly discussed in the next section.
 We would like to write the propagator in a form that may be decomposed
 into a product of fast and slow propagators. To this end we 
 demand that only the resonant regime of the propagator be properly
 recovered. In other words when $ (x-x^{'}) \approx \pm v_{F} (t-t^{'}) $. 
\[
\left< \psi^{\dagger}(x^{'}t^{'}) 
\mbox{        }\psi(xt)\right> 
= \frac{1}{2 \pi i} 
 \mbox{       } e^{ i k_{F}^2 (t-t^{'}) }
\left[ \frac{ e^{ i k_{F}\mbox{      }
[ (x-x^{'})  - v_{F} (t-t^{'}) ] } }
{ (x-x^{'}) - v_{F} (t-t^{'}) }
- \frac{  e^{ -i k_{F}\mbox{      }
 [(x-x^{'}) + v_{F} (t-t^{'}) ] } }
{ (x-x^{'}) + v_{F} (t-t^{'}) } \right]
\]
\begin{equation}
 = (2i) \mbox{       }\frac{(x-x^{'})}{\pi i} 
 \mbox{       } e^{ -i k_{F}^2 (t-t^{'}) }
\left[ \frac{ sin [ k_{F}(x-x^{'}) ] }
{ (x-x^{'})^2 - v^2_{F} (t-t^{'})^2 } \right]
\end{equation}
In other words,
\begin{equation}
\left< \Psi_{fast}^{*}(x^{'}t^{'})
 \mbox{      }\Psi_{fast}(xt)  \right>
 = \rho^{0} \mbox{        }e^{ -i k_{F}^2 (t-t^{'}) }
\mbox{       }  sin [ k_{F}|x-x^{'}| ] 
\end{equation}
Thus we may deduce,
\begin{equation}
\left< \psi_{slow}^{\dagger}(x^{'}t^{'})
 \mbox{      }\psi_{slow}(xt)  \right>
 = \frac{ |x-x^{'}| }{ k_{F} } \mbox{       } 
\left[ \frac{ 2 }{ (x-x^{'})^2 - v^2_{F} (t-t^{'})^2 } \right]
\end{equation}
In what follows we shall only insist on being able get the resonant
 part of the slow propagator correctly. In other words,
 $ [ (x-x^{'})^2 - v^2_{F} (t-t^{'})^2 ]^{-1} $. We shall not
 be too concerned if we fail to get the $ |x-x^{'}| $ in the numerator.
 From the hydrodynamic approach we have,
\begin{equation}
\left< \psi_{slow}^{\dagger}(x^{'}t^{'}) \psi_{slow}(xt) \right> 
= e^{ -\frac{1}{2} \sum_{ q \neq 0, n } 
\left( 2 - e^{ i q  \mbox{   }
(x-x^{'}) } e^{ z_{n} (t-t^{'}) } 
 - e^{ i q \mbox{     }
(x^{'}-x) } e^{ z_{n} (t^{'}-t) }  \right)
\mbox{      }F(q,n) }
\end{equation}
\[
F(q,n) = \left< \left( X_{ q, n } - G(q, n) \mbox{       }
\rho_{ -q, -n } \right)
\left( X_{ -q,-n } - G(-q,-n) \mbox{       }
\rho_{ q, n } \right) \right>
\]
\begin{equation}
 =  \left< X_{ q, n } X_{-q, -n } \right> 
 -  G(q, n) \mbox{       } \left< \rho_{ -q, -n } X_{-q,-n } \right>
 - G(-q,-n) \mbox{       } \left< X_{q,n } \rho_{q,n } \right>
 + G(q, n) \mbox{       }  G(-q,-n) \mbox{       }
  \left< \rho_{ q, n } \rho_{-q, -n } \right> 
\end{equation}
In one dimension we may deduce,
\begin{equation} 
\left< \rho_{ q, n } \rho_{ -q, -n } \right>
 = \frac{ 2 N^{0} }{ \beta } \left( \frac{ z^2_{n} }{ q^2 }
 + v_{F}^2 \right)^{-1}
\end{equation}
\begin{equation} 
\left< X_{ q, n } X_{ -q, -n } \right>
 = \frac{1}{2 \beta N^{0} } \left( \frac{ z_{n}^2 }{ v^2_{F} }
 + q^2 \right)^{-1}
\end{equation}
\begin{equation} 
\left< X_{ q, n } \rho_{ q, n } \right>
 = \left( - \frac{ z_{n} }{ \beta q^2 } \right)
\mbox{      } \left( \frac{ z^2_{n} }{ q^2 }
 + v_{F}^2 \right)^{-1}
\end{equation}
If we set,
\begin{equation} 
 G(q,n) = \frac{ 3 v^2_{F} }{ 2 N^{0} \mbox{     }z_{n} } 
\end{equation}
then,
\begin{equation} 
F(q,n) = \left[ \frac{ 2 v^2_{F} }{ \beta N^{0} }  \right]
  \mbox{       }  \left( z^2_{n} + v_{F}^2 q^2 \right)^{-1}
\end{equation}
 Here we have taken the liberty to ignore the pole at $ z_{n} = 0 $ since
 this may be cancelled by an appropriate time-independent
 additive contribution
 to $ \Lambda([\rho];{\bf{x}},t) $. This does not affect the action but
 serves to regularise the above integrals.
\begin{equation}
\left< \psi_{slow}^{\dagger}(x^{'}t^{'}) \psi_{slow}(xt) \right> 
= e^{ -\frac{1}{2} \sum_{ q \neq 0, n } 
\left( 2 - e^{ i q  \mbox{   }
(x-x^{'}) } e^{ z_{n} (t-t^{'}) } 
 - e^{ i q \mbox{     }
(x^{'}-x) } e^{ z_{n} (t^{'}-t) }  \right)
\mbox{      }F(q,n) }
\end{equation}

\newpage

\[
 \frac{1}{2} \sum_{ q \neq 0, n } 
\left( 2 - e^{ i q  \mbox{   }
(x-x^{'}) } e^{ z_{n} (t-t^{'}) } 
 - e^{ i q \mbox{     }
(x^{'}-x) } e^{ z_{n} (t^{'}-t) }  \right)
\mbox{      }F(q,n) 
\]
\begin{equation}
= \int^{ \infty }_{ 0 } \frac{ dq }{q} \mbox{     }
\left( 2 - e^{ i q  \mbox{   }
(x-x^{'}) } e^{ i \mbox{    }v_{F}|q| (t-t^{'}) } 
 - e^{ i q \mbox{     }
(x^{'}-x) } e^{ i\mbox{   }v_{F}|q| (t-t^{'}) }  \right)
\end{equation}
Thus we have,
\begin{equation}
\left< \psi_{slow}^{\dagger}(x^{'}t^{'}) \psi_{slow}(xt) \right> 
 \sim \left[ (x-x^{'})^2 - v^2_{F} (t-t^{'})^2 \right]^{-1}
\end{equation}
 Now we compute the total momentum-total momentum correlation function
 of the free Fermi theory.
 This will be important when we compute a.c. conductivity.
 To this end we write,
\begin{equation}
{\bf{P}}(t) = - \sum_{ {\bf{q}},  n, n^{'}  }
 ( i {\bf{q}} ) \mbox{        }  \rho_{ {\bf{q}}, n }
X_{ {\bf{q}}, n^{'} } \mbox{        }
 e^{ ( z_{ n^{'} } - z_{n} ) \mbox{       }t }
\end{equation}
In one dimension, in the hydrodynamic language we may write,
\[
< T \mbox{        }P(t) \mbox{   }P(t^{'}) > = 
 \sum_{ q,  n, n^{'}  }
\sum_{ q^{'},  m, m^{'}  }
 ( i q ) ( i q^{'} )  \mbox{        } 
\left<  \rho_{ q, n }
X_{ q, n^{'} } 
  \rho_{ q^{'}, m }
X_{ q^{'}, m^{'} } \right> \mbox{        }
 e^{ ( z_{ m^{'} } - z_{m} ) \mbox{       }t^{'} }
 e^{ ( z_{ n^{'} } - z_{n} ) \mbox{       }t }
\]
\[
=  \sum_{ q,  n, n^{'}  }
  \mbox{        } 
q^2  \mbox{        } 
 \left<  \rho_{ q, n }   \rho_{ -q, -n } \right>
\left< X_{ q, n^{'} } 
X_{ -q, -n^{'} } \right> \mbox{        }
 e^{ ( z_{ n^{'} } - z_{n} ) \mbox{       }(t-t^{'}) }
\]
\[
 -  \sum_{ q,  n, n^{'}  }
  \mbox{        } 
q^2  \mbox{        } 
 \left<  \rho_{ q, n } X_{ q, n } \right>
\left< X_{ q, n^{'} } \rho_{ q^{'}, n^{'} } \right>
\mbox{        }
 e^{ ( z_{ n^{'} } - z_{n} ) \mbox{       }(t-t^{'}) }
\]
\[
=  \sum_{ q,  n, m  }
  \mbox{        } 
q^4  \mbox{        } 
  \frac{ v^2_{F} }{ \beta^2 } \left(  z^2_{n} 
 + v_{F}^2 q^2 \right)^{-1} 
  \left( (z_{n} + z_{m})^2
 + v^2_{F} q^2 \right)^{-1}
\mbox{        }
 e^{ z_{ m } \mbox{       }(t-t^{'}) }
\]
\[
 - \sum_{ q, n, m  }
  \mbox{        } q^2 \mbox{         }
  \left( \frac{ z_{n} (z_{ n } + z_{m}) }{ \beta^2 } \right)
\mbox{      } \left( z^2_{n} 
 + v_{F}^2 q^2 \right)^{-1}
\mbox{      } \left( (z_{n}+z_{m})^2 
 + v_{F}^2 q^2 \right)^{-1} \mbox{        }
 e^{ z_{ m }  \mbox{       }(t-t^{'}) }
\]
\begin{equation}
 = \sum_{ q, m } 
|q|^3  \mbox{        } 
  \frac{ v_{F} }{ \beta }
\frac{1}{ z^2_{m} + 4 v^2_{F} q^2 } 
\mbox{        }
 e^{ z_{ m } \mbox{       }(t-t^{'}) }
 - \sum_{ q, m }
  |q|^3\mbox{        } \frac{ v_{F} }{ \beta } \mbox{         }
 \frac{1}{ z^2_{m} + 4 v^2_{F} q^2 } 
 \mbox{        }
 e^{ z_{ m }  \mbox{       }(t-t^{'}) } = 0
\end{equation}
In the Fermi language,
\begin{equation}
\left< T \mbox{       }P(t) \mbox{  }P(t^{'}) \right>
 = \sum_{ k,k^{'} } (k.k^{'}) \mbox{        }
\left< n_{k} n_{ k^{'} } \right> 
 = \sum_{k} k^2 \mbox{        }\theta(k_{F} - |k|)
 = N^{0} \mbox{     } \frac{ k^2_{F} }{3}
\end{equation}
 Thus there seems to be a discrepency. 
 We must then take the point of view that the hydrodynamic approach predicts
 the right time-dependent part of the total-momentum total
 momentum correlation function.

\newpage

\section{ The Luttinger Model }

The action for the Luttinger model reads as follows.
\[
S_{Lutt} = \sum_{ q, n } (-i \beta z_{n}) \rho_{ q, n } 
X_{ q, n } 
 + i \beta N^{0} \sum_{ q, n } q^2 \mbox{     } 
 X_{ q, n }  X_{ -q, -n } 
\]
\begin{equation}
 + \frac{ i \beta }{ 4 N^{0} }
\sum_{ q, n } q^2  
 \rho_{ q, n }
 \rho_{ -q, -n }
 + i \beta  \sum_{ q \neq 0 n }
 \frac{ v^2_{F} }{ 4 N^{0} }  \mbox{        }
\rho_{ q, n }
\rho_{ -q, -n } 
+ \frac{ i \beta \rho^{0} V_{0} }{ 2 N^{0} }  \sum_{ q \neq 0 n }
\rho_{ q, n } \rho_{-q,-n}
\end{equation}
Here $ V_{0} > 0 $ is a positive constant signifying repulsion.
Thus we may write $ v_{eff} = v_{F} 
\left( 1 + \frac{ V_{0} }{ \pi v_{F} } \right)^{\frac{1}{2}} $.
 Unfortunately it seems that we have to redefine $ G $
(we may euphemistically call this `renormalization'). Thus we 
 retry  $ G(q,n) = \lambda \frac{ v^2_{F} }{ N^{0} \mbox{     }z_{n} } $
\begin{equation}
F(q,n) = \left[ \frac{ v^2_{eff} }
{2 \beta N^{0} } 
 - \lambda \frac{ 2 v^2_{F} }{ \beta N^{0} }  \mbox{       }
 + 4\lambda^2 \frac{ v^4_{F}/v^2_{eff} }
{ 2 \beta N^{0} }
 \right] \mbox{     }\left( z^2_{n}
 + v_{eff}^2 q^2 \right)^{-1}
\end{equation}

\[
\frac{1}{2} \sum_{ q \neq 0, n } 
\left( 2 - e^{ i q  \mbox{   }
(x-x^{'}) } e^{ z_{n} (t-t^{'}) } 
 - e^{ i q \mbox{     }
(x^{'}-x) } e^{ z_{n} (t^{'}-t) }  \right)
\mbox{      }F(q,n) 
\]
\begin{equation}
= \int^{ \infty }_{0} \frac{ dq }{q} \mbox{     } 
\left( 2 - e^{ i q  \mbox{   }
(x-x^{'}) } e^{ i v_{eff} |q| (t-t^{'}) } 
 - e^{ i q \mbox{     }
(x^{'}-x) } e^{ i v_{eff} |q|  (t-t^{'}) }  \right)
\mbox{      }
\left[ \frac{ v_{eff} }
{4 v_{F} } 
 - \lambda \frac{ v_{F} }{v_{eff} }
  + \lambda^2 \frac{ v^3_{F} }{ v^3_{eff} }
 \right] 
\end{equation}
 It is clear that this approach will not give the right exponent since
 to get the right exponent we have to choose some complicated $ \lambda $.
 Thus we shall not insist on getting the one-particle properties right.

\section{Electron-Phonon System}

Consider the lagrangian for phonons.
\begin{equation}
L_{phonons} = \sum_{ {\bf{q}} } 
\frac{1}{4 \Omega_{ {\bf{q}} } } \mbox{    }
\left( \frac{ \partial u_{ {\bf{q}} }(t) }
{ \partial t }  \frac{ \partial u_{ -{\bf{q}} }(t) }
{ \partial t } 
 - \Omega^2_{ {\bf{q}} }  \mbox{       }
u_{ {\bf{q}} }(t) u_{ -{\bf{q}} }(t) \right)
\end{equation}
 Here $ u_{ {\bf{q}} } = b_{ {\bf{q}} } + b^{\dagger}_{ -{\bf{q}} } $
 is the phonon displacement and
 the phonon dispersion is chosen to be 
 acoustic : $ \Omega_{ {\bf{q}} } = v_{s} |{\bf{q}}| $. 
 From Mahan \cite{Mahan} we may write down
 the hamiltonian for the electron-phonon interaction neglecting
 umklapp processes,
\begin{equation}
H_{ep} = \frac{1}{ \sqrt{ \Omega } } \sum_{ {\bf{q}} \sigma }
M_{ {\bf{q}} } \mbox{       }\rho_{ {\bf{q}} \sigma }(t)
 \mbox{       }u_{ {\bf{q}} }(t)  
\end{equation}
\begin{equation}
M_{ {\bf{q}} } = - V_{ei}({\bf{q}}) \mbox{        }|{\bf{q}}| 
\mbox{       } \left( \frac{1}{ 2 M_{ion} \mbox{      }
\eta \mbox{   }v_{s} |{\bf{q}}| } 
\right)^{ \frac{1}{2} }
\end{equation}
We choose $ -V_{ei}({\bf{q}}) = Z^{eff}_{ion}  (4 \pi e^2) / {\bf{q}}^2 $.
The overall action may be written as,
\[
S = \sum_{ {\bf{q}} \sigma, n }
 (-i \beta z_{n}) \rho_{ {\bf{q}}\sigma, n } 
X_{ {\bf{q}} \sigma, n } 
 + \frac{ i \beta N^{0} }{2}
\sum_{ {\bf{q}} \sigma, n } {\bf{q}}^2 \mbox{       }
 X_{ {\bf{q}} \sigma, n }  X_{ -{\bf{q}} \sigma, -n } 
+ i \beta \sum_{ {\bf{q}} n } z_{n} \mbox{       }
C({\bf{q}}n) \mbox{       }
\rho_{ {\bf{q}} \sigma, n }
 \rho_{ -{\bf{q}} \sigma, -n }
\]
\begin{equation}
+ (i \beta) \sum_{ {\bf{q}} n } 
\frac{ \left( z_{n}^2 + \Omega^2_{ {\bf{q}} } \right) }
{ 4 \Omega_{ {\bf{q}} } }
\mbox{       }
u_{ {\bf{q}}, n } u_{ -{\bf{q}},-n } 
+ \frac{ i \beta }{ \sqrt{ \Omega } }
\sum_{ {\bf{q}} \sigma, n } M_{ {\bf{q}} } \rho_{ {\bf{q}} \sigma, n } 
u_{ {\bf{q}}, n }
\end{equation}
 Here $ \eta = N_{ion}/N_{e} = 1/Z_{ion} $ is the ratio of the number
 of ions to the number of electrons and $ \hbar = 2m_{e} = 1 $. 
 Also $ Z^{eff}_{ion} \sim 1 $ is the effective charge seen by the outer
 electrons. In the case of phonons the momentum transfer
 $ |{\bf{q}}| < \Lambda_{D} $, the Debye cutoff.
 We may integrate out the phonons first
 and write down the effective action for the electrons. 
\[
S_{eff} = \sum_{ {\bf{q}} \sigma,  n } 
(-i \beta z_{n}) \rho_{ {\bf{q}}\sigma, n } 
X_{ {\bf{q}}\sigma, n } 
 + i \beta \frac{ N^{0} }{2}
\sum_{ {\bf{q}}\sigma, n } {\bf{q}}^2 \mbox{       }
 X_{ {\bf{q}} \sigma,  n }  X_{ -{\bf{q}} \sigma, -n } 
 + i \beta \sum_{ {\bf{q}} n } z_{n} \mbox{       }
C({\bf{q}}n) \mbox{       }
\rho_{ {\bf{q}} \sigma, n }
 \rho_{ -{\bf{q}} \sigma, -n }
\]
\begin{equation}
- \frac{ i \beta }{ 2 N^{0}  }
\sum_{ {\bf{q}} \sigma, \sigma^{'}, n } 
 \frac{ 2 \rho^{0}
 M^2_{ {\bf{q}} } \Omega_{ {\bf{q}} }  }
{ z_{n}^2 + \Omega^2_{ {\bf{q}} } } \mbox{       }
 \rho_{ {\bf{q}} \sigma, n } 
 \rho_{ -{\bf{q}} \sigma^{'}, -n } 
\end{equation}
 In order to test for superconductivity we have to compute these
 quantities. The first is the one-particle 
 dynamical density of states which is
 related to the full dynamical propagator. The second
 is the momentum distribution which should not have
 a discontinuity at the Fermi surface.
 The third is Yang's off-diagonal long range order correlation function. 
 Unfortunately to get these right we have to get the one-particle propagator
 right. We shall now postulate that current algebra is more sacred than
 Fermi statistics and set (following the suggestion of A.H. Castro-Neto),
\begin{equation}
\psi_{slow}({\bf{x}}\sigma,t) \approx e^{ -i \mbox{      }
\Pi({\bf{x}}\sigma,t) } 
\mbox{       }
\sqrt{ \rho^{0} }
\end{equation}
\begin{equation}
{\bf{J}}({\bf{x}}\sigma,t) \approx 
- \rho^{0} \mbox{     }\nabla \Pi({\bf{x}}\sigma,t)
\end{equation}
In three dimensions we may write,
\[
\left< \rho_{ {\bf{q}} \sigma ,n }
\rho_{ -{\bf{q}} \sigma , -n } \right>_{0} =
\frac{1}{ -i \beta } \sum_{ {\bf{k}} } n_{\beta}({\bf{k}}+{\bf{q}}/2) 
( 1 - n_{\beta}({\bf{k}}-{\bf{q}}/2) )
\mbox{         }
\frac{ e^{ \beta \frac{ {\bf{k.q}} }{m}  }
 - 1 }
{  i \frac{ {\bf{k.q}} }{m} - z_{n}  }
\]
\[
 = \frac{ k^2_{F} }{ -i \beta } \frac{ V }{ (2\pi)^2 }
\int^{ \infty }_{ -\infty }   \mbox{       }dy \mbox{       }
 \int^{ +1 }_{ -1 } d \mbox{  }cos \theta \mbox{         }
 \left( \frac{1}{ 
e^{ \beta [ v_{F} y 
 + \frac{ v_{F}\mbox{    }q }{2} 
 \mbox{   }cos \theta ] } +1 } \right) \mbox{    }
\]
\[
\mbox{       }\left( \frac{1}{ e^{ \beta [ -v_{F} y 
 + \frac{ v_{F}\mbox{    }q }{2} \mbox{   }cos \theta ]  } +1 } \right)
\mbox{         }
\frac{ e^{ \beta v_{F} q \mbox{       }cos \theta }
 - 1 }{  i  v_{F} q \mbox{     }cos \theta  - z_{n}  }
\]
\begin{equation}
 =|{\bf{q}}| \left( \frac{ k^2_{F} }{ -i \beta } \right)
 \frac{ V }{ (2\pi)^2 }
 \int^{ 1 }_{ 0 } dx 
\mbox{         }
Tanh \left[ \beta \frac{ v_{F} q }{4}\mbox{       }x \right]
\frac{ x }{  i  v_{F} q \mbox{     }x + z_{n}  }
\end{equation}

\section{ Conclusions }

 It should be clear to the reader by now that bosonization is a powerful
 method that could hold the key to understanding a wide variety of 
 phenomena using an extremely general and economical set of rules.
 The sea-boson method has reduced what was till recently, difficult
 research problems into difficult homework problems.
 Now the hope is that the hydrodynamic approach will reduce everything
 to homework problems. The preferred method favored by condensed matter
 theorists involves starting from experiments
 and working backwards and trying to come up with some minimal description
 may sound as the most economical way of proceeding. However this approach
 is fraught with ambiguity and lacks the generality that the present theory
 has. In fact this approach of starting from experiments may be compared
 to this humorous situation. Imagine an athlete when asked to run a marathon
 stands on the finish line. When the umpire shouts `ready!', he takes a few
 steps back ...`get set'...`go!'. The athlete runs like the wind and declares
 himself a winner of the marathon after a fraction of a second. Such an athlete
 would be the laughing stock of the sporting world. Yet it is this approach 
 that most condensed matter theorists favor. The reader may argue that there
 is no telling where the starting line is, it could be 22 light years away
 (as in String Theory or M-theory) rather than 22 miles away. 
 While this is a valid objection to my stand in principle, in practice it
 is not. Rutherford and others have already showed us using the 
 starting-from-experiments approach that (for most purposes) 
  the atom is the building block of
  matter and the atom is made of positive
 and negatively charged point particles making the atom neutral
 and these obey quantum mechanics. Thus there is 
 no need to repeatedly ask the experimentalist for data. All that these 
 experiments in condensed matter will tell us
 is that matter is made of positive and
 negative charges and quantum mechanics is important.
 We already know that. We have to now work
 out the consequences of this knowledge.
   The main aim of this preprint is to convince the reader of the urgency
 of somewhow making this scheme work for then we will be able to
 compute the observable properties of a theory of neutral matter where the
 only adjustable parameters are the relative concentrations of the 
 various elements that make up the material.

\section{Appendix A}

 Here we find the most general solution to the recursion relation
 for the phase functional that determines the statistics. We find
 that it has to be in the specific form that has been used in the main
 text. Consider Eq.(~\ref{recur}) with $ m $
 being in general, a function 
 of the density and also a function of
 the pair of points $ {\bf{x}}, {\bf{x}}^{'} $
 in such a way that
 $ m([\rho];{\bf{x}},{\bf{x}}^{'}) = - m([\rho];{\bf{x}}^{'},{\bf{x}}) $.  
 We may formally Fourier transform the phase functional as follows.
\begin{equation}
\Phi([\rho];{\bf{x}}) = \int D[P] \mbox{       }
e^{ i P.\rho } \mbox{          }\phi([P];{\bf{x}}) 
\end{equation}
Therefore, Eq.(~\ref{recur}) may be rewritten as follows.
\begin{equation}
\phi([P];{\bf{x}}) \mbox{       }
( e^{ - i P({\bf{x}}^{'}) } - 1 )
 - \phi([P];{\bf{x}}^{'}) \mbox{       }
( e^{ - i P({\bf{x}}) } - 1 )
 = \pi \mbox{       }{\tilde{m}}([P];{\bf{x}},{\bf{x}}^{'})
\end{equation}
This means that $ {\tilde{m}} $ has to be of the form,
\begin{equation}
{\tilde{m}}([P];{\bf{x}},{\bf{x}}^{'}) = 
 ( e^{ - i P({\bf{x}}^{'}) } - 1 ) \mbox{       } ( e^{ - i P({\bf{x}}) } - 1 )
  \mbox{       } \left( F([P];{\bf{x}}) - F([P];{\bf{x}}^{'}) \right) 
\end{equation}
In other words,
\begin{equation}
\frac{ \phi([P];{\bf{x}}) }{  e^{ - i P({\bf{x}}) } - 1  }
 - \frac{ \phi([P];{\bf{x}}^{'}) }{  e^{ - i P({\bf{x}}^{'}) } - 1  }
 =  F([P];{\bf{x}}) - F([P];{\bf{x}}^{'})
\end{equation}
This means,
\begin{equation}
\frac{ \phi([P];{\bf{x}}) }{  e^{ - i P({\bf{x}}) } - 1  }
 =  F^{'}([P];{\bf{x}}) 
\end{equation}
Or,
\begin{equation}
 \Phi([\rho];{\bf{x}}) = 
 \Lambda([\{ \rho({\bf{y}}) - \delta^{d}({\bf{y}}-{\bf{x}}) \}];{\bf{x}}) 
 - \Lambda([\rho];{\bf{x}}) 
\end{equation}
 Unfortunately for current algebra to be strictly obeyed, we must have
 $ \Lambda([\rho];{\bf{x}})  \equiv  \Lambda([\rho])  $ independent of
 $ {\bf{x}} $. This violates the statitstics requirement. Thus we shall
 have to make do with recovering the current algebra as a limit. 
 
 Next we prove Eq.(~\ref{eqvr}). To do this we write,
\begin{equation}
v^{r}(x) = \int D[P] \mbox{        }e^{i P. \rho }
\mbox{          }{\tilde{v}}^{r}(x;P)
\end{equation}
 We plug this into the equation below 
\begin{equation}
[-i \partial_{x} \frac{ \delta }{ \delta \rho(x) }, v^{r}(x^{'})]
 = [-i \partial_{x^{'}} \frac{ \delta }{ \delta \rho(x^{'}) }, v^{r}(x)]
\end{equation}
to obtain,
\begin{equation}
[-i \partial_{x} \frac{ \delta }{ \delta \rho(x) },
  \int D[P] \mbox{        }e^{i P. \rho }
\mbox{          }{\tilde{v}}^{r}(x^{'};P)]
 = [-i \partial_{x^{'}} \frac{ \delta }{ \delta \rho(x^{'}) }, 
 \int D[P] \mbox{        }e^{i P. \rho }
\mbox{          }{\tilde{v}}^{r}(x;P)] 
\end{equation}

\begin{equation}
 \int D[P] \mbox{        }e^{i P. \rho }
\mbox{          }\left[ \partial_{x} P(x) \mbox{        }
{\tilde{v}}^{r}(x^{'};P)
 - \partial_{x^{'} } P(x^{'}) \mbox{        }
{\tilde{v}}^{r}(x;P) \right]
 = 0
\end{equation}
Or,
\begin{equation}
\partial_{x} P(x) \mbox{        }
{\tilde{v}}^{r}(x^{'};P)
 - \partial_{  x^{'} } P(x^{'}) \mbox{        }
{\tilde{v}}^{r}(x;P) = 0
\end{equation}
In other words,
\begin{equation}
{\tilde{v}}^{r}(x;P) =  G([P]) \mbox{        }\partial_{x} P(x)
\end{equation}
and  Eq.(~\ref{eqvr}) follows.

\section{Appendix B}

 Here we would like to ascertain whether or not the velocity operator
 of fermions is irrotational. To this end we note that if,
\begin{equation}
{\bf{J}} = - \rho \nabla \Pi
\end{equation}
then in three dimensions,
\begin{equation}
\rho \nabla \times {\bf{J}} = \nabla \rho \times {\bf{J}}
\end{equation}
Thus we would like to verify whether or not this is obeyed.
In general we may write,
\begin{equation}
{\bf{J}}({\bf{x}}) = \frac{1}{V} \sum_{ {\bf{k}} {\bf{q}} }
{\bf{k}} c^{\dagger}_{ {\bf{k}} + {\bf{q}}/2 }
c_{ {\bf{k}} - {\bf{q}}/2 } e^{-i {\bf{q.x}} }
\end{equation}
\begin{equation}
\rho({\bf{x}}) = \frac{1}{V} \sum_{ {\bf{k}} {\bf{q}} }
 c^{\dagger}_{ {\bf{k}} + {\bf{q}}/2 }
c_{ {\bf{k}} - {\bf{q}}/2 } e^{-i {\bf{q.x}} }
\end{equation}
From this we can see that,
\begin{equation}
\rho \nabla \times {\bf{J}} - \nabla \rho \times {\bf{J}}
 = -i\frac{1}{ V^2 } \sum_{ {\bf{k}} {\bf{q}} }
\sum_{ {\bf{k}}^{'} {\bf{q}}^{'} }
\left[ ({\bf{q}}-{\bf{q}}^{'}) \times {\bf{k}} \right] \mbox{         }
c^{\dagger}_{ {\bf{k}}^{'} + {\bf{q}}^{'}/2 }
c_{ {\bf{k}}^{'} - {\bf{q}}^{'}/2 }
c^{\dagger}_{ {\bf{k}} + {\bf{q}}/2 }
c_{ {\bf{k}} - {\bf{q}}/2 } \mbox{           }
e^{-i ({\bf{q}}+{\bf{q}}^{'}).{\bf{x}} }
\label{RHOFUN}
\end{equation}
 This operator is not identically zero but is zero in suitably
 restricted Hilbert space.  We have assumed in the text that 
 this is the case even though it is strictly speaking not right.
 Besides, current algebra is obeyed only if the velocity
 operator is strictly irrotational as already shown.

 Consider the hamiltonian,
\begin{equation}
H = \sum_{ {\bf{q}} \neq 0 } N^{0} \epsilon_{ {\bf{q}} }
X_{ {\bf{q}} }X_{ -{\bf{q}} }
 + \sum_{ {\bf{q}} \neq 0 } \frac{ \epsilon_{ {\bf{q}} } }{4 N^{0} }
\rho_{ {\bf{q}} } \rho_{ -{\bf{q}} }
\label{HXRHO}
\end{equation}
where $ [X_{ {\bf{q}} }, \rho_{ {\bf{q}} }] = i $ and all other commutators
involving any two of these is zero. There seems to be only one inequivalent way
in which we may diagonalise Eq.(~\ref{HXRHO}). It is to decompose
 Eq.(~\ref{HXRHO}) into oscillators with just one momentum
 label $ b_{ {\bf{q}} } $. This means we may write,
\begin{equation}
X_{ {\bf{q}} } = \frac{ i }{ 2  \sqrt{ N^{0} } }
\left( b_{ -{\bf{q}} } - b^{\dagger}_{ {\bf{q}} } \right)
\end{equation}
\begin{equation}
\rho_{ {\bf{q}} } = \sqrt{ N^{0} }
 \left( b_{ {\bf{q}} } + b^{\dagger}_{ -{\bf{q}} } \right)
\end{equation}
\[
H = \sum_{ {\bf{q}} \neq 0 } N^{0} \epsilon_{ {\bf{q}} }
 (\frac{ i }{ 2  \sqrt{ N^{0} } })^2
\left( b_{ -{\bf{q}} } - b^{\dagger}_{ {\bf{q}} } \right)
\left( b_{ {\bf{q}} } - b^{\dagger}_{ -{\bf{q}} } \right)
 + \sum_{ {\bf{q}} \neq 0 } \frac{ \epsilon_{ {\bf{q}} } }{4 N^{0} }
  N^{0} \mbox{      }
 \left( b_{ {\bf{q}} } + b^{\dagger}_{ -{\bf{q}} } \right)
  \left( b_{ -{\bf{q}} } + b^{\dagger}_{ {\bf{q}} } \right)
\]
\begin{equation}
= \sum_{ {\bf{q}} \neq 0 }  \epsilon_{ {\bf{q}} } 
b^{\dagger}_{ {\bf{q}} }b_{ {\bf{q}} }
\end{equation}
 This decomposition describes bosons. To describe fermions we need a
 dispersion of the kind $ v_{F} |q| $ rather than
 $ \epsilon_{q} = q^2/(2m) $. This is impossible to achieve with the
 hamiltonian in Eq.(~\ref{HXRHO}). This fact can be verfied independently
 using the equation of motion approach that also suggests that
 the dispersion of the modes is $ \epsilon_{ {\bf{q}} } $. 
 Thus we need an additional phase functional $ \Lambda $ to give
 us a linear dispersion.


\begin{thebibliography}{25}

\bibitem[1]{Landau} L.D. Landau, {\it{Collected Papers}}, 
1965, Pergamon Press, Oxford.

\bibitem[2]{Sunakawa} S. Sunakawa, S. Yamasaki and T. Kebukawa, Prog. Theor. Phys. {\bf{38}}, 804 (1967); {\bf{41}}, 919 (1969); {\bf{44}} 565 (1970); {\bf{49}} 1802 (1973).

\bibitem[3]{Raja}  A. K. Rajagopal and G. S. Grest, Phys Rev A
10, 1837 (1974); Phys. Rev. A10, 1395 (1974);
 Prog. Theor. Phys. 52, 811 (1974); {\bf 52}, 1719 (1974) Errata.;
A.K. Rajagopal in Quantum Fluids, in Particular, Superfluid Helium, 
Three Lectures in the Bose Symposiums on Statistical Physics (July 
15-27, 1974).  Supplement to Journal of Ind. Inst. of Sci. 85 (1975).

\bibitem[4]{Sharp} G. A. Goldin, R. Menikoff and D. H. Sharp, J. Math. Phys.
 {\bf{21}} (1980); R. Menikoff, D. H. Sharp, J. Math. Phys. {\bf{18}}, 471
 (1977); R.F. Dashen and D. H. Sharp, Phys. Rev. {\bf{165}}, 1857 (1968).

\bibitem[5]{Jackiw} R. Jackiw and A.P. Polychronakos, Phys. Rev. D,
 {\bf{62}} 085019 (2000) ; B. Bistrovic, R. Jackiw, H. Li, V. P. Nair
 and S.-Y. Pi,  Phys. Rev. D, {\bf{67}} 025013 (2003).

\bibitem[6]{Setlur1}G.S. Setlur and Y.C. Chang, Phys. Rev. B, {\bf{57}},
 15144 (1998); G.S. Setlur and D.S. Citrin, Phys. Rev. B, {\bf{65}},
 165111 (2002).

\bibitem[7]{preprint} G.S. Setlur, X-ray Edge Spectra from Sea-bosons,
cond-mat/0111423.

\bibitem[8]{Mahan}  G.D. Mahan, {\it{Many Particle Physics}},
 Plenum Press, \copyright 1990.

\bibitem[9]{Mattis} D.C. Mattis and E.H. Lieb,  J. Math. Phys. {\bf{6}},
                                           304 (1965).

\bibitem[10]{Bose} U(1) Gauge Theory as Quantum Hydrodynamics,
\\ G.S. Setlur, cond-mat/0210673.

\bibitem[11]{Kadanoff}L. Kadanoff and G. Baym,
 {\it{ Quantum Statistical Mechanics }}, Benjamin, New York.

\end{thebibliography}
\end{document}